\documentclass[12pt]{article}
\input epsf
\usepackage{epsfig}
\begin{document}
\def\C{{\mathbf{C}}}
\def\D{{{D}}}
\def\PT{{{PT}}}
\def\mod{{\mathrm{mod}\,}}
\def\Ree{{\mathrm{Re}\,}}
\def\R{{\mathbf{R}}}
\title{High energy
eigenfunctions of one-dimensional Schr\"odinger operators
with polynomial potentials}
\author{A. Eremenko\thanks{Supported by NSF grants
DMS-0555279 and DMS-0244547.},$\;$ A. Gabrielov and B. Shapiro} 
\date{\today}
\maketitle

\begin{abstract}
For a class of one-dimensional Schr\"odinger operators
with polynomial potentials that includes Hermitian and
$\PT$-symmetric operators, we show that the
zeros of scaled eigenfunctions have a 
limit distribution
 in the complex plane as the eigenvalues
tend to infinity. This limit distribution
depends only on the degree of the polynomial potential
and on the boundary conditions. 
\vspace{.1in}

\end{abstract}

\noindent
{\bf 1. Introduction}
\vspace{.1in}

We begin with an eigenvalue problem of the form
\begin{equation}
\label{1}
-y^{\prime\prime}+P(x)y=\lambda y,\quad y(-\infty)=y(\infty)=0,
\end{equation}
where $P(x)=x^d+\ldots$ is a real monic polynomial of even
degree $d$. The boundary condition is equivalent to
$y\in L^2(\R)$. It is well-known that the spectrum
of this problem is discrete, all eigenvalues are real and simple,
and they can be arranged into an increasing sequence
$\lambda_0<\lambda_1<\ldots\to+\infty$. Moreover,
$$\displaystyle
\lambda_n\sim\left(\frac{\pi dn}{2B(3/2,1/d)}\right)^{
\frac{\scriptstyle 2d}{\scriptstyle d+2}}=
\left(\frac{\sqrt{\pi}\Gamma(3/2+1/d)n}{\Gamma(1+1/d)}
\right)^{\frac{\scriptstyle 2d}{ \scriptstyle d+2}},
\quad n\to\infty,
$$
where $B$ and $\Gamma$ are the Euler's functions. 
A general reference for these facts is~\cite{Titch}.

Eigenfunctions $y_n$ are entire functions of order
$(d+2)/2$, and in this paper we study the distribution of 
their zeros in the complex plane when $n$ is large.
When $d=2$ and $P$ is even, we have $y_n(x)=H_n(x)\exp(-x^2/2),$ where $H_n$
are the Hermite
polynomials, and asymptotic distribution of zeros is known
in this case in great detail \cite{Szego}.
The case $d=4$, 
which is called sometimes an anharmonic oscillator
or a double well potential, was studied much,
but most attention was payed to the properties of
eigenvalues, rather than the properties of eigenfunctions.

In \cite{EGS}, we proved that for $d=4$ and even $P$,
all zeros of all eigenfunctions belong to the union
 of the real
and imaginary axis. The main result of the present paper
implies that after an appropriate rescaling of the independent
variable, zeros of eigenfunctions have a limit distribution
in the plane,
which depends only on $d$. This will be derived from the 
asymptotics of $\log|Y_n|$ as $n\to\infty$,
where $Y_n$ is a properly rescaled eigenfunction.

Let us consider real zeros first. According to the
Sturm--Liouville theory, $y_n$ has exactly $n$ real zeros,
all of them simple. A classical argument shows that all
these real zeros belong to the interval $(a_n^-,a_n^+)$
where $a_n^-$ and $a_n^+$ are the smallest and the largest
roots of the equation $P(x)-\lambda_n=0$. The asymptotics
$$|a_n^{\pm}|\sim \lambda_n^{1/d}$$
suggests to define the rescaling:
\begin{equation}
\label{rescaling}
Y_n(z)=y(\lambda_n^{1/d}z).
\end{equation}
Now we denote by $\nu_n$ the counting measure of
the roots of $Y_n$ (for every set $X\subset\C$,
$\nu_n(X)$ is the number of roots
on $Y_n$ in $E$). Our main result, Theorem~2 below,
has the following
corollary:
$$\frac{\nu_n}{n}\to c_d\sqrt{1-x^d}\, dx,\quad -1\leq x\leq 1,$$
where the convergence is the usual weak convergence of measures:
$\mu_n\to\mu$ means that $\int\phi d\mu_n\to\int\phi d\mu$
for each continuous function with compact support.
The normalizing constant is
$$\displaystyle c_d=\frac{2}{d}B(3/2,1/d).$$
For $d=2$ the limit distribution is the ``semi-circle law'',
the well-known asymptotic distribution of zeros of Hermite's
polynomials. 

A theorem of Hille \cite[Theorem 11.3.3]{Hille} implies that for
every $r\in(0,1)$ there exists $n_0(r)>0$ such that for all
$n>n_0$, all zeros of $Y_n$ in the disc $\{ z:|z|\leq r\}$
are real.

Zelditch \cite{Zelditch} studies Laplace--Beltrami
eigenfunctions on real analytic manifolds.
Under certain conditions on the manifold, he extends
the eigenfunctions to a complex neighborhood of
the manifold and obtains a limit distribution of
their zeros. Asymptotics in
the complex domain help to study the distribution
of real zeros. The same happens in our case.

Passing to the consideration of complex zeros, we make
the same rescaling (\ref{rescaling}), and consider
the rescaled counting measures 
\begin{equation}
\label{mun}
\mu_n=\nu_n/n
\end{equation}
of zeros
of $Y_n$ in the complex plane.

Our main result
will imply that these measures $\mu$ converge weakly 
to an explicitly described
limit measure, which depends only on $d$.

Our results also apply to the so-called $\PT$-symmetric
Schr\"odinger operators which were intensively studied in
the recent years \cite{Be2,Shin}. Let $P$ be a complex polynomial of degree $d$
with the
property $P(-\overline{z})=\overline{P(z)}$. Schr\"odinger operators
with such potential $P$ are called $\PT$-symmetric.
Every $\PT$-symmetric potential can be written as
$P(z)=P_1(iz)$, where $P_1$ is a polynomial with real coefficients.
A real potential $P$ is $\PT$-symmetric if and only if
$P$ is even.

Following K. Shin \cite{Shin}, we consider the generalized
eigenvalue problem which contains both self-adjoint
and $\PT$-symmetric cases.
\begin{equation}
\label{3}
-y^{\prime\prime}+Py=\lambda y,
\end{equation}
with
$$P(z)=(-1)^\ell(iz)^d+\sum_{k=1}^{d-1}a_{k}z^k,$$
where the coefficients $a_k$ are arbitrary complex
numbers, and the boundary condition is
\begin{equation}\label{4}
y(re^{i\theta})\to 0,\quad r\to\infty\quad
\mbox{for}\quad\theta=-\frac{\pi}{2}\pm
\frac{(\ell+1)\pi}{d+2},
\end{equation}
where $1\leq \ell\leq d-1$.
The self-adjoint problem (\ref{1}) corresponds to the case that
$d$ is even, $\ell=d/2$, and $a_k$ are real.
The usual boundary condition imposed on a $\PT$-symmetric
potential corresponds to $\ell=1$ \cite{Be,Be2}.

The following result belongs to Sibuya \cite{Sibuya}
and K. Shin \cite{Shin}.
\vspace{.1in}

{\bf Theorem A.} {\em The spectrum of the problem $(\ref{3}),\;
(\ref{4})$ is discrete, and to each eigenvalue
corresponds a one-dimensional space of eigenfunctions.
Eigenvalues satisfy}
\begin{equation}
\label{5}
\lambda_n\sim\left(\frac{\sqrt{\pi}\Gamma(3/2+1/d)n}{\sin(\ell\pi/d)
\Gamma(1+1/d)}\right)^{\frac{\scriptstyle 2d}{\scriptstyle d+2}},\quad n\to\infty.
\end{equation}
So we see from the asymptotics that the eigenvalues are
``asymptotically real'' in the sense that their arguments
tend to zero. Shin proved that in the $\PT$-symmetric case
(that is when $a_k=(ib_k)^k$ with real $b_k$), all eigenvalues
but finitely many are actually real.  

Our main result, will imply 
that zeros of eigenfunctions $y_n$ of the problem (\ref{3}),
(\ref{4}), 
when properly rescaled as
in (\ref{rescaling}) have a limit distribution that depends only
on $d$ and $\ell$. The support of the limit distribution consists
of some Stokes lines (which we later define precisely),
and our result can be considered as a rigorous confirmation
of the results of numerical computations of Bender,
Boettcher and Savage \cite{Be}.

Functions $Y_n$ satisfy the differential equations of the form
\begin{equation}\label{asde}
Y_n^{\prime\prime}=\lambda_n^{2/d}(P(\lambda_n^{1/d}z)-\lambda_n)Y_n
=
k_n^2((-1)^\ell(iz)^d-1+o(1))Y_n,
\end{equation}
where
$k_n=\lambda_n^{(d+2)/(2d)}\to\infty,$
as $n\to\infty$. Here we choose the branch of $\lambda^{(d+2)/(2d)}$
which is positive on the positive ray.

It is well-known (see, for example \cite{Evgr,Head})
that the asymptotic behavior, as $n\to\infty$,
of solutions of such differential
equations depends on the quadratic differential 
\begin{equation}\label{Qd}
Q_{d,\ell}(z)dz^2=((-1)^\ell (iz)^d-1)dz^2.
\end{equation}
We recall the terminology and some
 known facts about such quadratic
differentials. For the general theory of quadratic
differentials we refer to \cite{Jenkins,Strebel}, and
for applications to differential equations to
\cite{Evgr,Fedor}. 

Let $Q$ be an arbitrary polynomial.
The zeros of $Q$ are called the {\em turning points}.
The curves on which $Q(z)dz^2<0$ are called the
{\em (vertical) trajectories}.
Each branch of the integral $\int\sqrt{Q}dz$
maps trajectories into vertical lines.
The trajectories make a foliation of the plane
with the singularities at the turning points.
The leaves of this foliation are maximal smooth open curves
on which $Q(z)dz^2<0$ holds. Each leaf is homeomorphic to
an open interval, and each end of a leaf is either at infinity
or at a turning point. Those leaves which have at least one end
at a turning point are called the {\em Stokes lines}.
A Stokes line whose both ends are turning points is called
{\em short}. At each simple zero of $Q$ exactly three
Stokes lines converge, and they make equal angles of $2\pi/3$
between them. 

The components of the complement of the union of turning points
and Stokes lines will be called the {\em Stokes regions}.
These regions are unbounded and simply connected.
The closure of each Stokes region (in $\C$) is homeomorphic
either to a closed half-plane or to a closed strip.
We say that these regions are {\em of the half-plane type}
or {\em of the strip type}, respectively.

Turning points, Stokes lines and Stokes regions make
a decomposition of the plane which we call the {\em Stokes complex}.

Our first result is the topological description of the Stokes
complex corresponding to the differential (\ref{Qd}).
\vspace{.1in}

\noindent
{\bf Theorem 1}. {\em 
The Stokes complex
of $Q_{d,\ell}dz^2$ is symmetric with respect to the 
imaginary axis and has the following property:
Every turning point $v$ which does not lie
on the imaginary axis is connected by a short Stokes
line with the turning point $-\overline{v}$ symmetric
to $v$ with respect to the imaginary axis.}
\vspace{.1in}

It is easy to see that this theorem yields a complete
topological description of the Stokes complex.
The turning points are the roots of the equation
$(-1)^\ell (iz)^d=1$. All these roots are simple,
so three Stokes lines meet at each turning point.

If $v$ is a turning point in the (open) right half-plane,
then there is one short Stokes line from $v$ to $-\overline{v}$.
The other two Stokes lines originating at $v$ go to infinity.
Indeed, if one of those two were short, we would obtain a
bounded Stokes region which is impossible.
These two Stokes lines are contained in the right half-plane
(otherwise they would intersect the symmetric Stokes lines
originating at $-\overline{v}$ which is impossible.)
The picture in the left half-plane is symmetric to the picture in
the right half-plane.

If $v$ belongs to the imaginary axis, that is $v=\pm i$,
then all three Stokes
lines originating from $v$ are unbounded. One of them coincides
with a ray of the imaginary axis $(\pm i,\pm i\infty)$
and the other two are symmetric with respect to the imaginary axis.

See Figures 1--9 
where the Stokes complexes of $Q_{d,\ell}dz^2$
for $d=2,3,4$ and $6$ and
various values of $\ell$ are
represented.
\begin{center}
\vfill
\epsfxsize=2.0in%
\centerline{\epsffile{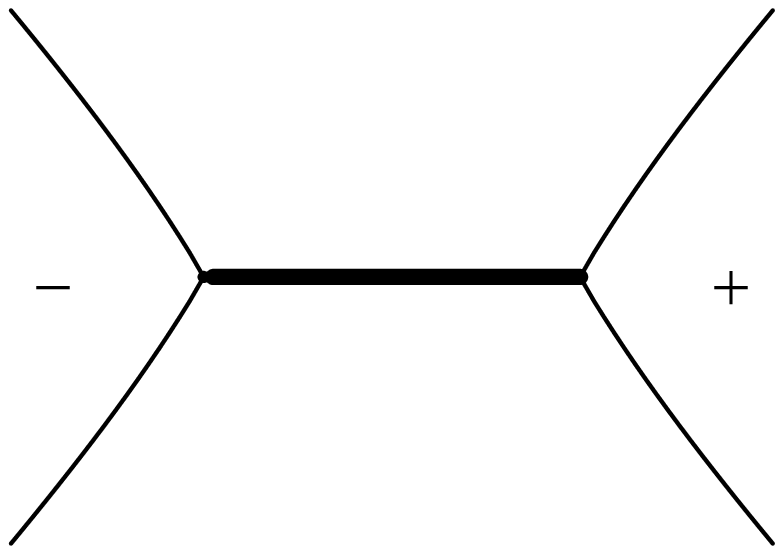}}
\nopagebreak
Fig. 1. $d=2,\;\ell=1$.
\vfill

\centerline{\epsffile{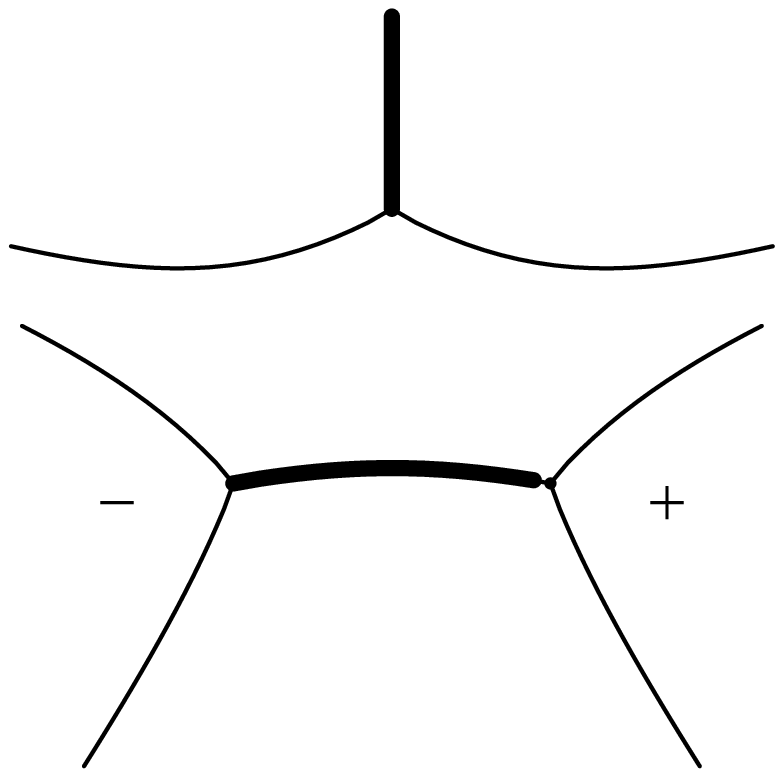}}
\nopagebreak
Fig. 2. $d=3,\;\ell=1$.
\vfill

\newpage
\centerline{\epsffile{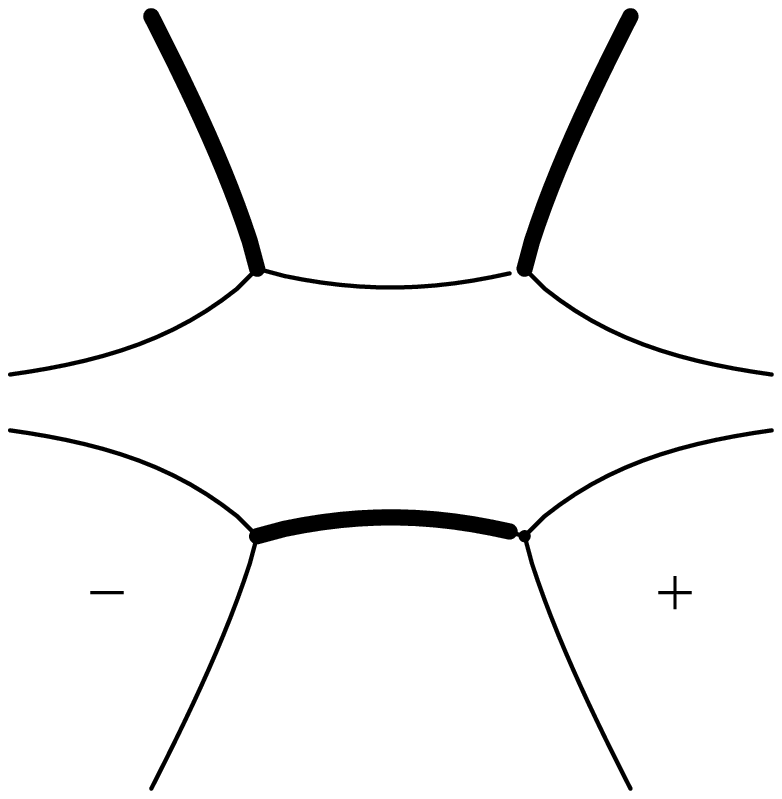}}
\nopagebreak
Fig. 3. $d=4,\;\ell=1$.
\vfill

\centerline{\epsffile{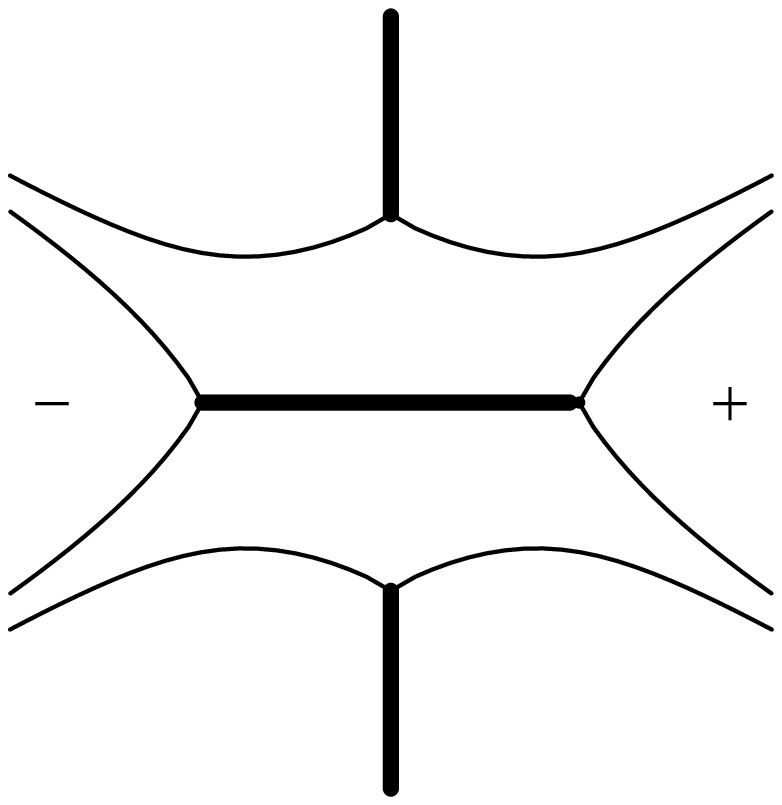}}
\nopagebreak
Fig. 4. $d=4,\;\ell=2$.
\vfill

\centerline{\epsffile{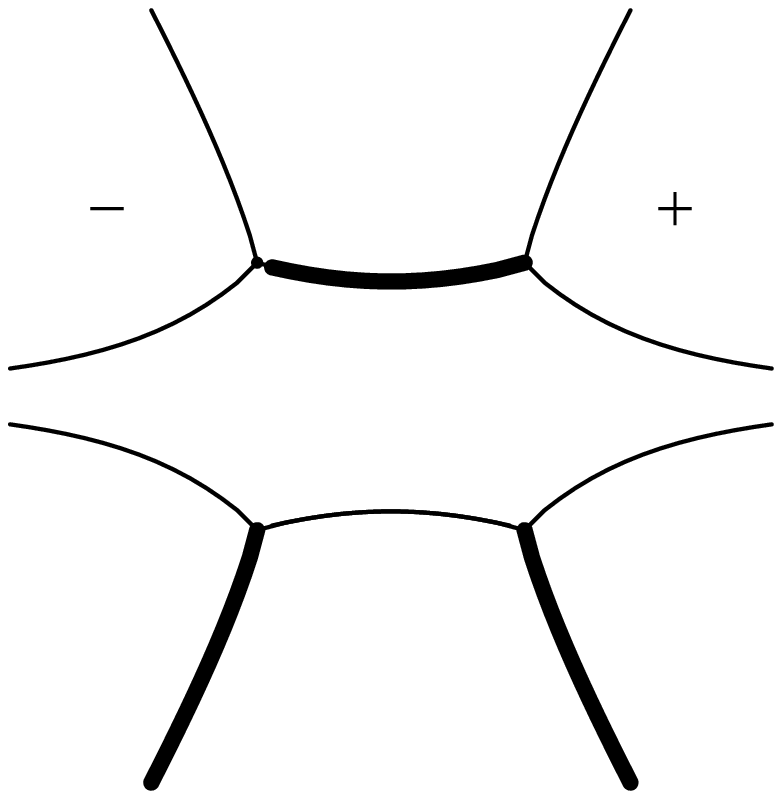}}
\nopagebreak
Fig. 5. $d=4,\;\ell=3$.
\vfill

\newpage
\centerline{\epsffile{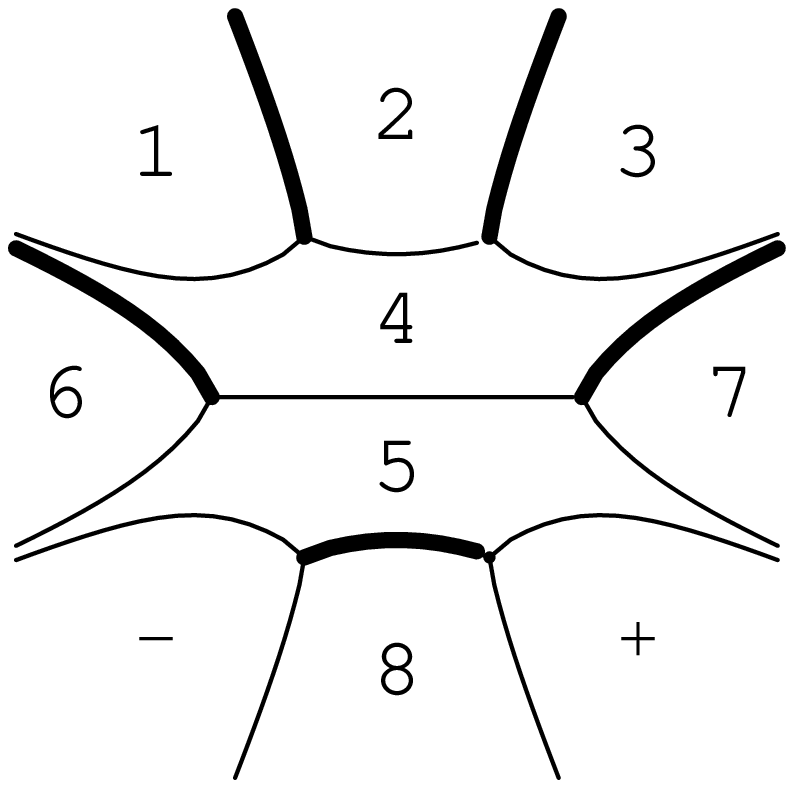}}
\nopagebreak
Fig. 6. $d=6,\;\ell=1$.
\vfill

\centerline{\epsffile{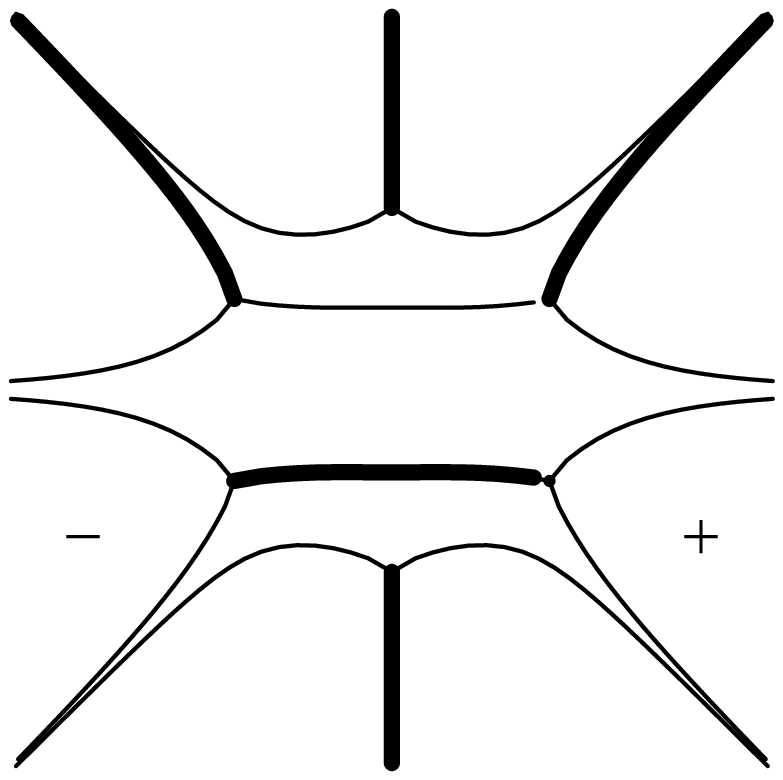}}
\nopagebreak
Fig. 7. $d=6,\;\ell=2$.
\vfill

\centerline{\epsffile{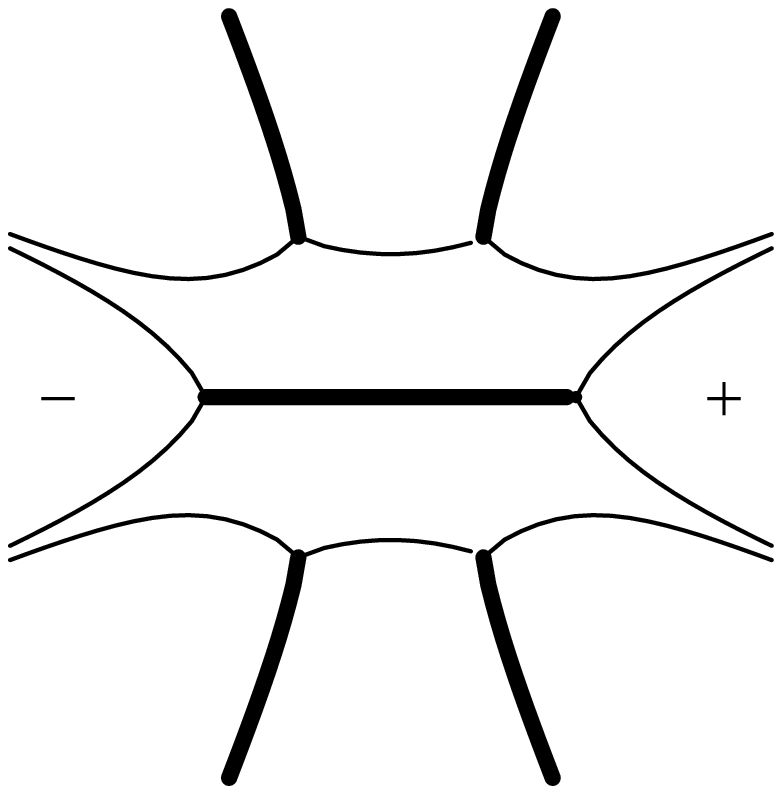}}
\nopagebreak
Fig. 8. $d=6,\;\ell=3$.
\vfill

\newpage
\centerline{\epsffile{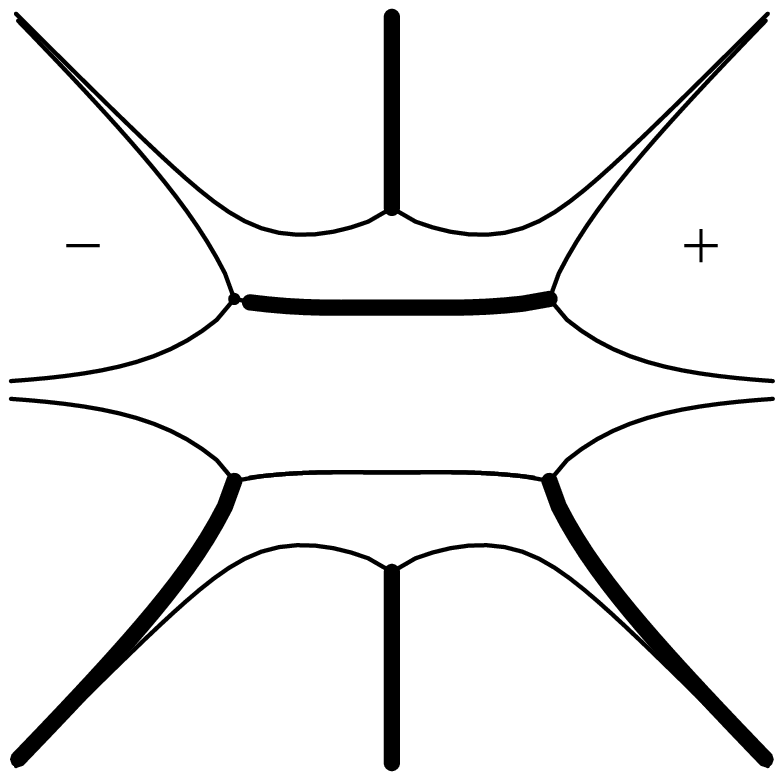}}
\nopagebreak
Fig. 9. $d=6,\;\ell=4$.
\end{center}
\vspace{.2in}

It is easy to see that each Stokes complex has exactly $d+2$
Stokes regions of the half-plane type. These regions are asymptotic
to the sectors which are bounded by the {\em Stokes directions}
$$\{ e^{i\theta}: \Ree\int_0^{\exp(i\theta)}\sqrt{
(-1)^\ell (it)^d}dt=0\}\quad\mbox{that is}\quad
\theta=-\frac{\pi}{2}+\frac{\pi(l+2k)}{d+2}.$$
The bisectors of the angles between adjacent Stokes directions
are called the {\em anti-Stokes directions}.
Each of the $d+2$ anti-Stokes directions approximately bisects
some Stokes region of the half-plane type.
Thus the boundary condition (\ref{4}) indicates that $y(z)\to 0$
as $z\to\infty$ on two anti-Stokes directions symmetric
with respect to the imaginary axis. This boundary condition 
singles out two Stokes regions of the type of half-plane:
the right region $\omega^+$ and the left region $\omega^-$,
so that the eigenfunction tends to zero along the bisectors
of $\omega^-$ and $\omega^+$. In Figures 1--9
the regions $omega^+$ and $\omega^-$ are marked
by $+$ and $-$ signs, respectively.
Figures 1, 4 and 8 correspond to self-adjoint problems. 

It follows from the topological description
of the Stokes complex of $Q_{d,\ell}dz^2$ above
that the closures of $\omega^+$ and $\omega^-$
are disjoint.

To state our main result, we need to define certain Stokes lines
in the Stokes complex of $Q_{d,\ell}dz^2$
which we call {\em exceptional}. There will be exactly one
exceptional Stokes line originating at each turning point.

Let $v^+$ and $v^-$ be the turning points
on the boundaries of $\omega^+$ and $\omega^-$, respectively.
Then $v^+$ and $v^-$ do not belong to the imaginary axis,
so Theorem~1 implies that there is a short Stokes line
$E_0=(v^-,v^+)$. This short Stokes line will be exceptional.
For example, for the self-adjoint problem (\ref{1})
we have $E_0=(-1,1)$.

If $v=i$ or $v=-i$ is a turning point, then
the Stokes line $(i,+\infty i)$ or $(-i,-\infty i)$
is exceptional.

For the rest of turning points, exceptional Stokes lines are
defined as follows.
The complement of the
set $\overline{\omega^+\cup\omega^-}\cup E_0$,
where the bar stands for the closure, consists
of two components, we call these components $D^+$ (containing
the positive imaginary ray)
and $D^-$ (containing the negative imaginary ray).
Let $v$ be a turning point in $D^+$ and not on the imaginary axis.
Let $L'$ and $L^{\prime\prime}$ be the two unbounded Stokes lines
originating at $v$. Of these two Stokes lines we choose that one
which lies between the other one and the positive imaginary axis,
and call this chosen line exceptional. (Any family
of disjoint curves tending to infinity in $D^+$
can be linearly ordered, for example anticlockwise; we
used the word ``between'' in the sense of this order).
Similarly, if $v$ is a turning point in $D^-$ and not on the imaginary
axis, then of the two unbounded Stokes lines originating from $v$,
the exceptional one is that which lies between the other one
and the negative imaginary axis.

Exceptional Stokes lines
are shown as bold lines in our figures.

Let $E$ be the union of all exceptional Stokes lines and
all turning points. We call $E$ the exceptional set and
denote $$\Omega=\C\backslash E.$$ 
Then $\Omega$ is a doubly connected region,
and $\sqrt{(-1)^\ell(iz)^d-1}$
has
two single-valued holomorphic branches in $\Omega$.
Consider the harmonic function
in $\Omega$
\begin{equation}
\label{A}
u(z)=\Ree\int_0^z\sqrt{(-1)^\ell (it)^d-1}dt=\Ree\int_0^z
\sqrt{Q_{d,\ell}(t)}dt.
\end{equation}
Here we choose the branch of the $\sqrt{Q_{d,\ell}}$ in such a way
that $u(z)\to-\infty$ as $z\to\infty$
along the anti-Stokes directions in $\omega^+$ and $\omega^-$.
(The normalization (\ref{4}) and the top coefficient
of the potential $P$ in (\ref{3}) are chosen to make
the choice of such branch of $\sqrt{Q_{d,\ell}}$ possible.)

The integral in the definition of $u$
has one period corresponding to loops
in $\Omega$ around the short Stokes line $E_0=(v^-,v^+)$,
but this period is pure imaginary,
because $Q_{d,\ell}dz^2<0$ on $E_0$,
thus the real part of this integral is a well defined harmonic
function in $\Omega$. In fact $u$ is continuous
and subharmonic in the whole plane
(that the jumps of the integral
on the exceptional Stokes lines is pure imaginary can be 
seen from the very definition of the Stokes lines).
Now we can state our main result.
\vspace{.1in}

\noindent
{\bf Theorem 2.} {\em Let $y_n$ be an eigenfunction
of the problem $(\ref{3}), (\ref{4})$ corresponding to
the eigenvalue $\lambda_n$, and normalized so that 
$|y_n(0)|+|y_n^\prime(0)|=1.$
Put $Y_n(z)=y_n(\lambda_n^{1/d}z)$.
Then
$$\frac{1}{n}\log|Y_n|\to c_{d,\ell} u(z),\quad n\to\infty,$$
uniformly on compact subsets of $\Omega$, and also
in the sense of Schwartz distributions $\D'$ in $\C$.
Here $u$ is the function defined in $(\ref{A})$, and 
$$c_{d,\ell}=\frac{\sqrt{\pi}\Gamma(3/2+1/d)}{\sin(\ell\pi/d)\Gamma(1+1/d)}.$$}
\vspace{.1in}

Uniform convergence implies that the functions
$Y_n$ have no zeros
on a compact $K\subset\Omega$ when $n>n_0(K)$.
This can be made more precise by replacing a compact set by an
unbounded subset of $\Omega$ which we describe below as
an ``admissible set''. Each compact subset of $\Omega$ is contained
is some admissible set.

The Laplace operator is continuous in $\D'$, and the Riesz
measure $\mu=(2\pi)^{-1}\Delta u$
can be easily explicitly computed:
\vspace{.1in}

\noindent
{\bf Corollary.} {\em The normalized 
counting measures $\mu_n=\nu_n/n$ of
the zeros of $Y_n$ converge weakly to the measure $$\mu
=c_{d,\ell}\sqrt{|Q_{d,\ell}(z)}|dz|,$$
supported on $E$.}
\vspace{.1in}

In particular, in the self-adjoint case (\ref{1})
the limit density on $(-1,1)$ is given by the formula
$c_d\sqrt{1-x^d}$, as advertised. Paper \cite{Be}
contains nice pictures
showing the zeros of the rescaled eigenfunctions
of $\PT$-symmetric operators clustering around $E_0$.

\vspace{.1in}

\noindent
{\bf 2. Proofs}
\vspace{.1in}

{\em Proof of Theorem 1.}
It is convenient to make the change of the independent
variable $z\mapsto iz$
which reduces our quadratic differential to the form
$(\pm z^d+1)dz^2$. It is clear that the Stokes complex of
this new differential is symmetric with respect to the real line.
All we have to prove is that there is a short Stokes line
from every turning point in the upper half-plane
to the complex conjugate turning point.
The proof is performed
in several steps.
\vspace{.1in}

{\em Step 1.} Let $P(z)=(z^d+1)$. We prove that
the Stokes complex of $P(z)dz^2$ contains a short
Stokes line between the turning points $v_1=e^{i\pi/d}$
and $v_{-1}=\overline{v_1}=e^{-i\pi/d}$.
To find the directions of the Stokes lines originating
at $v_1$, we write $P(z)=dv_1^{d-1}(z-v_1)+o(z-v_1),$ and
$$\int_{v_1}^z\sqrt{P(t)}dt=\frac{2}{3}\sqrt{d}
v_1^{(d-1)/2}(z-v_1)^{3/2}+o(z-v_1)^{3/2}.$$
So on the tangents to the Stokes lines meeting at $v_1$
we have $(z-v_1)^3v_1^{d-1}<0$.
Since $v_1^{d-1}=-1/v_1$, this is equivalent to
\begin{equation}
\label{one}
(z-v_1)^3\in v_1\R_+.
\end{equation}
Consider the sector
$S$ bounded by the segments $[0,1]$, $[0,v_1]$
and the arc of the unit circle $[1,v_1]$.
Our polynomial $P$ maps $S$ into the
first quadrant, thus for $z\in S$,
we have $\arg(z^d+1)\in(0,\pi/2)$.
This means that the direction $\phi$ of the vertical foliation
$P(z)dz^2<0$ in $S$ satisfies 
\begin{equation}
\label{two}
\pi/4<\phi+\pi k<\pi/2.
\end{equation}
There is one Stokes line originating at $v_1$ with the
direction
in this interval. According to (\ref{one}),
its direction at $v_1$ is $\pi/(3d)-2\pi/3$.
Because of the limitations (\ref{two}) on the direction of
this line in $S$,
it can leave $S$ only through
the interval $(0,1)$. On this interval it has to meet 
the symmetric Stokes line from $v_{-1}$.
Thus $v_1$ and $v_{-1}$ are
connected with a short Stokes line.
\vspace{.1in}

{\em Step 2.} 
The vertical foliation of the differential
$P_\theta(z)dz^2=e^{-2\theta}P(e^{-\theta} z)dz^2$
is obtained from
the vertical foliation of $P(z)dz^2$ by counterclockwise
rotation by $\theta$. In particular, the Stokes complex
of $P_\theta(z)dz^2$ contains a short
Stokes line between the
adjacent turning points $e^{i(\theta\pm\pi/d)}$.
\vspace{.1in}

{\em Step 3.} We prove that every
turning point $v_k=e^{\pi i(2k-1)/d}$
of $P(z)dz^2$ in the (open)
first quadrant
is connected to the symmetric turning point $\overline{v_k}$
by a short Stokes line. This we prove by induction
in $k$. For $k=1$ the statement was proved in Step 1.
Suppose it holds for $k\leq m-1$, and let $L_{m-1}$
be the Stokes line from $v_{m-1}$ to its conjugate.

The tangents to the Stokes lines originating at $v_m$
satisfy the equation $(z-v_m)^3\in v_m\R_+$
similar to (\ref{one}). In particular there
is one Stokes line $L_m$ starting at $v_m$ in the
direction $\phi_m=\pi(2m+1)/(3d)-2\pi/3$, so that
$-2\pi/3<\phi_m<-\pi/2$.

{}From Step 2 applied with $\theta=2\pi m/d$,
there is a Stokes line $U_m$ for the differential $P_\theta
dz^2$ connecting $v_m$ and $v_{m-1}$. Its tangent at 
$v_m$ has direction
$$\psi_m=
\phi_1+\theta=\pi/d-2\pi/3+2\pi m/d=
\pi(2m+1)/d-2\pi/3>\phi_m.$$
We are going to show that
$L_m$ never intersects $U_m$ in
the open unit disc. Proving this by contradiction,
we suppose that $z$ is a point of
intersection, and denote by $L_m^\prime$
and $U_m^\prime$ the pieces of $L_m$ and $U_m$ from $v_m$ to $z$.
Then the integrals 
$$\int_{L_m^\prime}\sqrt{P(z)}dz
\quad\mbox{and}\quad 
\int_{U_m^\prime}\sqrt{P_\theta(z)}dz$$
are both non-zero (because the imaginary part of
such integral is monotone along a Stokes line) and both
pure imaginary. But $P_\theta=e^{-2\pi im/d}P$ for our choice
of $\theta$, so we conclude that
$$\int_{L_m^\prime}\sqrt{P(z)}dz\neq 
\int_{U_m^\prime}\sqrt{P(z)}dz,$$
as these integrals have different arguments and are
non-zero.
This is impossible
because the integral of $\sqrt{P}dz$ should
be zero over any closed curve in the unit disc.
Thus $L_m$ and $U_m$
do not intersect in the open unit disc. 

For all $z$ in the unit disc, 
the direction of the vertical foliation
of $P(z)dz^2$ belongs to
the interval $\pi/4<\phi+\pi k<3\pi/4$. Hence
$L_m$ cannot leave the upper half of the unit
disc through the arc $[v_m,-1]$ of the unit circle.

Moreover, $L_m$ cannot intersect 
the Stokes line $L_{m-1}$ which exists by
the induction assumption.

This leaves only one possibility that $L_m$ leaves the
upper half of the unit disc through the real axis.
Then, by symmetry, it should continue to $v_{-m}$.
\vspace{.1in}

{\em Step 4.} Now we prove the same for the differential
$P_-(z)dz^2=(-z^d+1)dz^2$, namely that every turning point in
the first quadrant is connected by a short Stokes line with
its conjugate. First we prove that
the Stokes line $L_1$ starting at the turning
point $z_1=e^{2\pi i/d}$ in the direction $2\pi/(3d)-2\pi/3$
cannot intersect the Stokes line of $P_{\pi/d}dz^2=e^{-2\pi/d}
P_-(z)dz^2$
connecting $z_1$ with $0$, hence the only
possibility for $L_1$ to leave the upper half
of the unit disc is through the real line, so it should 
proceed to $\overline{z_{1}}$ by the symmetry.
Then we proceed by induction on $m$ to prove that
there are Stokes lines for $P_-dz^2$ connecting
the turning points $z_m$ in the first quadrant with
their complex conjugates.
The induction step is similar to the one in Step 3
and we leave it to the reader.
\vspace{.1in}

{\em Step 5.} That every turning point in the left half-plane
is connected by a short
Stokes line with its complex conjugate
is proved by the change of the independent variable
$z\mapsto -z$.

This completes the proof of Theorem 1.

\vspace{.2in}

Proceeding to the proof of Theorem~2
we begin with some rough estimates.
\vspace{.1in}

\noindent
{\bf Lemma 1.} {\em Consider a normalized solution of the
differential equation
$$y^{\prime\prime}=h^2Qy, \quad y(0)=y_0,\quad y'(0)=y_1,$$
where $h>0$ and $Q$ is a holomorphic function
satisfying $|Q(z)|\leq M$ for $|z|\leq R$.
Then we have
$$|y(z)|\leq\max\{ |y_0|,|y_1|\}\exp(hMR).$$
}
\vspace{.1in}

{\em Proof.}
We put $w_1=hy$ and $w_2=y'$, to obtain a matrix equation
$${\mathbf{w}'}(z)=A(z){\mathbf{w}}(z),$$ where
$${\mathbf{w}}=\left(\begin{array}{l}
 w_1\\ w_2\end{array}\right)\quad\mbox{and}\quad
A=
\left(\begin{array}{cc} 0&h\\hQ&0\end{array}\right).
$$
This implies
\def\w{{\mathbf{w}}}
$$\| \w(z)\|\leq\| \w(0)\|+\int_0^z\| A(t)\|\|\w(t)\|dt,$$
where we use the $\sup$-norms
$$\|\w\|=\max\{ |w_1|, |w_2|\},$$
and
$$\| A\|=\max\{ |a_{11}|+|a_{12}|,\; |a_{21}|+|a_{22}|\}\leq hM.$$
Applying Gronwall's lemma \cite[Thm. 1.6.6]{Hille}, we obtain
the result. 
\vspace{.1in}

Applying this result to the equation (\ref{asde}), and taking
into account the asymptotics of the eigenvalues
from Theorem~A,
we conclude that if the sequences
$(1/n)\log|Y_n(z_0)|$ and $(1/n)\log|Y_n^\prime(z_0)|$
are bounded at some point $z_0\in\C$, then
in every disc $|z|<R$ the
family of subharmonic functions 
\begin{equation}
\label{fam}
u_n=\frac{1}{n}\log|Y_n|
\end{equation}
is
uniformly bounded from above.
By a well-known theorem
(see, for example, \cite[Theorem 4.1.9]{Hor})
it follows that from every sequence of these functions
one can select a subsequence which either converges
uniformly on every disc to $-\infty$,
or converges in $D'(\C)$
(or in $L^1_{{\rm loc}}(\C)$ which is equivalent for such families
of subharmonic functions).
Convergence to $-\infty$ is excluded by normalization
of our eigenfunctions.

Thus the limit functions of our family are subharmonic,
and the same theorem \cite[Theorem 4.1.9]{Hor} says that the
Riesz measures of $u_n$ weakly converge to the
Riesz measures of $u$.

Thus all we need to do is to identify the
possible limit functions $u$, and to show that there
is only one.
This will be done by proving the uniform convergence part
of Theorem~2. Notice that subharmonic functions
are upper semi-continuous, so the restriction of $u$ on
$\Omega$ defines $u$ uniquely in the whole plane. 

To prove the uniform convergence
in Theorem~2, we 
need a finite open covering of $\Omega$
with the so-called canonical regions \cite{Evgr}.

Consider a quadratic differential $Q(z)dz^2$
with arbitrary polynomial $Q$.
The multi-valued function
\begin{equation}
\label{zeta}
\zeta(z)=\int^z\sqrt{Q(t)}dt
\end{equation}
has holomorphic branches in each Stokes region.
These branches are defined up to post-composition
with a transformation $z\mapsto\pm z+c$,
where $c$ is an arbitrary constant.
Each such branch maps its Stokes region univalently
onto some right or left half-plane, or onto a vertical
strip, depending on the type of the region \cite{Evgr,Fedor}.

A region $D$, which is a union of Stokes regions
and Stokes lines is
called a {\em canonical region} if there is a holomorphic branch
of $\zeta$ in $D$ that maps $D$ onto the plane with finitely
many slits along some vertical rays. 

It is clear that a canonical region contains exactly two
distinct Stokes regions of the half-plane type.
It is easy to state a criterion for a pair of Stokes regions
of the half-plane type to belong to some canonical region:
\vspace{.1in}

{\em Stokes regions $D_1$ and $D_2$ belong to a canonical region
if and only if there is a curve from $D_1$ to $D_2$ that
does not pass through the turning points, intersects
the Stokes lines transversally, and
intersects each component of the $1$-skeleton of the
Stokes complex at most once.}
\vspace{.1in}

For example, in Figure~6 there is a canonical region
containing the Stokes regions $-,5,6$, another
one containing $-, 5,4,1$ 
and so on, but there is no canonical
region containing $-$ and $+$.

It is easy to conclude from Theorem 1
that in the case that $Q=Q_{d,\ell}$, for each Stokes region
$D$, other than $\omega^-$, there is
a canonical region containing
$\omega^+$ and $D$. Similar statement holds with $\omega^-$
and $\omega^+$ interchanged. 
Moreover, for $Q=Q_{d,\ell}$, canonical regions containing $\omega^+$
cover $\Omega$, and canonical regions containing $\omega^-$
cover $\Omega$ as well. So the statement about uniform convergence
in Theorem~2 is enough to prove for all canonical regions
containing $\omega^+$ or~$\omega^-$.

Let $Q(z)dz^2$ be a quadratic differential with
a polynomial $Q$, and let a number $s>0$ be given.
A
smooth curve $\gamma:(-1,1)\to \C$
will be called $s$-admissible
if it has the following properties:
\vspace{.1in}

\noindent
(i) $\gamma(t)\to\infty,$ as $|t|\to 1$, 
\vspace{.1in}

\noindent
(ii) For every $t\in(-1,1)$, the (smaller)
angle between $\gamma'(t)$
and the vertical trajectory at $\gamma(t)$
is at least $s$.  
\vspace{.1in}

\noindent
(iii) For every $t\in(-1,1)$, the distance from
$\gamma(t)$ to the set of turning points is at least $s$.
\vspace{.1in}

A subset $K\subset \Omega$, will be called $s$-admissible,
if $K$ is a union of $s$-admissible
curves. 

A subset $K\subset\Omega$ or a curve $\gamma$ in $\Omega$
will
be called simply {\em admissible} if it is 
$s$-admissible for some $s>0$.
It is easy to see that interiors of admissible subsets
of canonical regions containing $\omega^+$ or $\omega^-$
cover $\Omega$. Two Stokes regions of the half-plane type
belong to some
canonical region if and only if there is an admissible
curve passing through these two Stokes regions.

Now we have to study the behavior of the Stokes complex
under small perturbations of the quadratic differential.
For our purposes the following lemma will be sufficient.
\vspace{.1in}

\noindent
{\bf Lemma 2.}
{\em Let $Q(z,h)$ be a polynomial of degree $d$ with
respect to $z$ 
whose coefficients are continuous functions of $h$, for $h$
in a neighborhood of $0$,
and suppose that $\deg_zQ(z,0)=d$, and
that $K$ is an admissible subset of some
canonical region of $Q(z,0)dz^2$.
Then there exists $h_0>0$, such that for $|h|<h_0$,
the differential $Q(z,h)dz^2$ has a canonical region in which
$K$ is admissible.}
\vspace{.1in}

{\em Proof}. Let $\gamma$ be an admissible curve
for $Q(z,0)dz^2$.
This means that $\gamma$ does not pass through
the zeros of $Q(z,0)$ and that
the angles between the tangent lines
to $\gamma$ and the vertical trajectories of $Q(z,0)dz^2$
are bounded from below by some $s>0$.
It is clear that for every compact piece of $\gamma$
these conditions persist for the vertical trajectories
of $Q(z,h)dz^2$ for small $h$.
Thus one only has to consider a
neighborhood of $\infty$. Suppose that
$Q(z,h)=a(h)z^d+\ldots$.
Then there exist $h_0>0$ and $R>0$ such that
$$|\arg Q(z,h)-d\arg z-\arg a(0)|<s/2,$$
whenever $|z|>R$ and $|h|<h_0$.
The angle between $\gamma'$ and the vertical
direction is
$$\arg\gamma'-\frac{\pi}{2}+\frac{\arg Q}{2}+\pi k.$$
so if this angle is at least $s$ for $h=0$, then it
is at least $s/2$ for $|h|<h_0$ and $|z|>R$.
This proves the lemma.
\vspace{.1in}

The following lemma is essentially well-known,
though we could not find
a convenient reference
in the literature, so we include a proof.
This lemma constitutes the essence of the Liouville method,
which is known to physicists as the WKB method
(see \cite{Head,Olver} for the general discussion
of the method, including history).
The closest statement
to ours is the one in \cite{Evgr},
but our proof is different,
and it is based on Liouville's transformation.\footnote{We
do not share the opinion of Fedoryuk
\cite[Ch. II, \S 1,
sect. 6]{Fedor}
that the Liouville's transformation is ``less convenient''
for complex functions than for the real ones.}
\vspace{.2in}

\noindent
{\bf Lemma 3.}
{\em Let $Q$ be an arbitrary polynomial, and consider
the differential equation
\begin{equation}
\label{7}
-y^{\prime\prime}+h^2Q(z)y=0,
\end{equation}
where $h>0$ is a parameter.
Let $D$ be a canonical region
for $Q(z)dz^2$, containing a Stokes
region $D_1$ of the half-plane type.
Let $\zeta=\Phi(z)$ be a branch of
$$\int_{z_0}^z\sqrt{Q(w)}dw$$
which maps $D$ onto the plane with vertical slits, and such that
$D_1$ corresponds to a left half-plane, and let $K\subset D$
be an
$s$-admissible set.
Then, for each $h>h_0(s,Q)$,
there exists a solution $y^*_h(z)$
of the differential equation
$(\ref{7})$, which satisfies
\begin{equation}
\label{kria}
y_h^*(z)=Q^{-1/4}(z)\exp(h\Phi(z))(1+\epsilon(z,h)),
\end{equation}
where
\begin{equation}
\label{esm}
|\epsilon(z,h)|\leq \frac{h_0}{h-h_0},\quad z\in K.
\end{equation}
Moreover, one can take
\begin{equation}
\label{h0}
h_0(s,Q)=\sup\int_\gamma\left|\frac{5}{16}\frac{{Q'}^2}{Q^3}-
\frac{Q^{\prime\prime}}{4Q^2}\right||d\zeta|,
\end{equation}
where the $\sup$ is over all $s$-admissible curves in $K$,
and derivatives in $(\ref{h0})$ are with respect to $z$.}
\vspace{.1in}

{\em Remark.}
It is easy to see that the integral in (\ref{h0})
is absolutely convergent. Assuming that the
leading coefficient of $Q$ has absolute value $1$
and all roots of $Q$ belong to the disc $|z|<2$,
we estimate this integral in terms of $s$.
We have
$|d\zeta|=\sqrt{|Q|}|dz|,$ and thus the integrand is at most
$A(s)(|\zeta|+1)^{-(d+4)/(d+2)}$ on any $s$-admissible
curve. 
Now, an $s$-admissible curve can be parametrized as
$\Phi^{-1}(\xi+if(\xi)):-\infty<\xi<\infty$,
where $f$ is a function whose derivative
satisfies $|f'(\xi)|\leq C(s)$.
So the integral does not exceed
$$A(s)\int_{-\infty}^\infty(|\xi|+1)^{-(d+4)/(d+2)}
\sqrt{1+C^2(s)}d\xi.$$
So,
the $\sup$ of these integrals defining $h_0$
is bounded by a constant which depends only on $s$,
for all $Q$ in a neighborhood of $Q_{d,\ell}$.
\vspace{.1in}

{\em Proof of Lemma 3}.
We use the change of the variables which is due to
Liouville (see, for example, \cite{Head}):
$$w(\zeta)=Q^{1/4}(\Phi^{-1}(\zeta))y(z(\zeta)),$$
Then the differential equation (\ref{7}) becomes
\begin{equation}
\label{8}
w^{\prime\prime}=hw+gw,
\end{equation}
where 
$$g(\zeta)=\left(-\frac{5}{16}\frac{{Q'}^2}{Q^3}+
\frac{Q^{\prime\prime}}{4Q^2}\right)\circ\Phi^{-1}(\zeta),$$
where the prime stands for the differentiation
with respect to $z$.

Now we consider the integral equation
\begin{equation}
\label{equa}
w(\zeta)=e^{h\zeta}+\frac{1}{2h}\int_{-\infty}^\zeta(e^{h(\zeta-t)}
-e^{h(-\zeta+t)})g(t)w(t)dt,
\end{equation}
where the path of integration $\gamma$ is a part of
the $\Phi$-image
of an $s$-admissible 
curve in $D$.
It is easy to verify directly
that every analytic solution of this integral
equation satisfies (\ref{8}). (To derive this integral equation
one ``solves'' (\ref{8}) by the method of variation of constants,
considering $gw$ as a given function, see~\cite{Head}).

Putting in (\ref{equa}) $w=e^{h\zeta}W$,
we obtain
\begin{equation}
\label{equa2}
W(\zeta)=1+\frac{1}{2h}\int_{-\infty}^\zeta (1-e^{2h(t-\zeta)})g(t)
W(t)dt.
\end{equation}
This we solve by the method of successive approximation:
set $W_0=0$ and define $W_{n+1}=1+F(W_n)$, where
$F$ is the integral operator in the right hand side of (\ref{equa2}).
Then we have for the sup-norms on
$\gamma$:
$$\| W_{n+1}-W_n\|_\gamma\leq\frac{1}{2h}\max_{t\in\gamma}
(1+
|e^{2h(t-\zeta)}|)\,\| W_n-W_{n-1}\|_\gamma 
\int_\gamma|g(t)|\,|dt|.$$
Property (ii) of admissible curves
implies that $\gamma$ intersects every vertical
line at most once, so $\Ree(t-\zeta)\leq 0$ on $\gamma$, and
$$|1+e^{2h(t-\zeta)}|\leq 2,$$
Thus for $h>h_0$, the
$W_n$'s converge geometrically, uniformly
on $\Phi(K)$ to a solution $W$ of the integral
equation (\ref{equa2}). We have
$$W=W_1+(W_2-W_1)+(W_3-W_2)+\ldots=1+\epsilon,$$
where $\epsilon$ satisfies (\ref{esm}).
This proves the lemma.
\vspace{.1in}

{\em Completion of the proof of Theorem 2}.
It remains to prove the statement about the
uniform convergence.
As admissible subsets
of canonical regions cover
$\Omega$, it is enough to prove that
the uniform convergence holds on each
admissible subset. Let $K$ be an $s$-admissible subset
of a canonical region $D$ containing $\omega^+$
 or $\omega^-$. Our functions
$Y_n$ satisfy differential equations (\ref{asde}).
For every sufficiently large $n$,
we set
$h_n=|k_n|$ and $$Q^*_n(z)=\exp(2i\arg k_n)
(P(\lambda_n^{1/d}z)-\lambda_n),
$$
so that 
\begin{equation}
\label{eqq}Y_n^{\prime\prime}=h_n^2Q^*_nY_n.
\end{equation}
In view of (\ref{asde}) and $\arg k_n\to 0$, we conclude
that $Q^*_n\to Q_{d,\ell}$ as $n\to\infty$ and
all conditions of Lemma~2 are satisfied.
When $n$ is large enough, we apply Lemma~2
to $Q^*_n$ and conclude that there exists a canonical region
for $Q^*_ndz^2$ which contains $K$.
According to the Remark after Lemma~3,
there is a bound for $h_0$ in Lemma~3 that is independent
of $n$. Applying Lemma~3 to this canonical
region and equation (\ref{eqq}), we obtain
a solution 
$y^*_n$ of the equation (\ref{eqq}) which satisfies
(\ref{kria}) with $h=h_n$ and some branch $\Phi_n$
of 
$$\int^z\sqrt{Q^*_n(t)}dt.$$
In particular, this solution $y^*_n$
tends to zero on the anti-Stokes direction
in $\omega^+$ or $\omega^-$.
Thus it is proportional to the eigenfunction. Our
normalization of the eigenfunction implies that
the coefficient of
proportionality satisfies $\log|c_n|=O(n)$.
So we obtain that
$$\frac{1}{h_n}\log|Y_n|\to u(z)+c,$$
uniformly on $K$, where $c$ may depend on $K$.
As the limit function exists and
is continuous in the whole plane, and $u$ is single-valued,
we conclude
that this last relation holds
in the whole plane in the sense of $D'$ 
with the same constant $c$.
Comparing the values near $0$ we conclude that $c=0$.

This completes the proof.
\vspace{.2in}

{\em Remark}. When we were completing this paper,
a preprint of Hezari was published \cite{Hezari}
where similar methods are used to study the
self-adjoint eigenvalue problem
$$y^{\prime\prime}=\lambda^2q(x)y,$$
with a real polynomial $q$, $q(x)\to +\infty$ as
$|x|\to\infty$ and
with the usual boundary conditions on the real axis.
Hezari shows that possible limit distributions of zeros
of eigenfunctions $y_n$ are supported by some Stokes lines
similar to our ``exceptional set''.
However, in his setting, a single limit
distribution as $n\to 0$
might not exist, as he shows by an example.

The main reason why the limit distribution exists in
our case is Theorem~1, which describes the topology
of a very special Stokes complex, while in Hezari's work
an arbitrary Stokes complex symmetric with respect to
the real line is involved.

{\em A. E.: Department of Mathematics

Purdue University

West Lafayette, IN 47907 USA

eremenko@math.purdue.edu
\vspace{.1in}

A. G.: Department of Mathematics

Purdue University

West Lafayette, IN 47907 USA

agabriel@math.purdue.edu
\vspace{.1in}

B. S.: Department of Mathematics

Stockholm University 

Stockholm, S-10691, Sweden

shapiro@math.su.de}

\begin{thebibliography}{1}
\bibitem{Be} C. Bender, S. Boettcher and V. M. Savage,
Conjecture on the interlacing of zeros in complex
Sturm-Liouville problems. 
J. Math. Phys. 41 (2000), no. 9, 6381--6387.
\bibitem{Be2} C. Bender,
Jun Hua Chen and K. Milton, 
$\PT$-symmetric versus
Hermitian formulations of quantum mechanics.
J. Phys. A 39 (2006), no. 7, 1657--1668.
\bibitem{EGS} A. Eremenko, A. Gabrielov and B. Shapiro,
Zeros of eigenfunctions of some anharmonic oscillators,
arXiv:math-ph/0612039.
\bibitem{Evgr} M. Evgrafov and Fedoryuk,
Asymptotic behaviour as $\lambda\to\infty$
of the solution of the equation $w^{\prime\prime}(z)
-p(z,\lambda)w(z)=0$ in the complex plane,
Russian Math. Surveys, 21 (1966) 1--48.
\bibitem{Fedor} M. Fedoryuk, Asymptotic analysis.
Linear ordinary differential equations,
Springer, Berlin, 1993.
\bibitem{Fedor2} M. Fedoryuk, Asymptotic expansions, in:
Analysis I, Encyclopaedia of Math. Sci.
 Springer, NY 1993.
\bibitem{Head} J. Heading, An introduction to the 
phase-integral methods, John Wiley, NY, 1962.
\bibitem{Hezari} H. Hezari,
Complex zeros of eigenfunctions of $1D$ Schr\"odinger
operators, arXiv:math-ph/0703028, 8 March 2007.
\bibitem{Hille} E. Hille, Ordinary differential equations
in the complex domain, J. Willey and Sons, NY, 1976.
\bibitem{Hor} L. H\"ormander,
Analysis of partial differential
operators I. Distribution theory and Fourier analysis,
Springer, Berlin, 1983.
\bibitem{Jenkins} J. Jenkins and D. Spencer,
Hyperelliptic trajectories, Ann. Math., 51 (1951), 4--35. 
\bibitem{Olver} F. Olver, Asymptotics and special functions,
Academic Press, London, 1974.
\bibitem{Shin} K. Shin, Schr\"odinger
type eigenvalue problems
with polynomial potentials: asymptotics of eigenvalues,
arXiv:math.SP/0411143, 7 Nov 2004.
\bibitem{Sibuya} Y. Sibuya, Global theory of a second order
linear ordinary differential equation with a polynomial
coefficient, North-Holland, NY, 1975.
\bibitem{Strebel} K. Strebel, Quadratic differentials,
Springer, Berlin, 1984.
\bibitem{Szego} G. Szeg\H{o}, Orthogonal polynomials,
AMS, Providence, RI, 1967
\bibitem{Titch} E. Titchmarsh, Eigenfunctions expansions
associated with second-order differential equations,
vol. 1, Oxford, Clarendon Press, 1946.
\bibitem{Zelditch} S. Zelditch,
Complex zeros of real ergodic eigenfunctions,
arXiv:math.SP/0505513, 8 Aug 2005.
\end{thebibliography}
\end{document}